\documentclass{article}
\usepackage{PRIMEarxiv}

\usepackage[T1]{fontenc}
\usepackage[utf8]{inputenc}
\usepackage{amsfonts}
\usepackage{booktabs}
\usepackage{nicefrac}
\usepackage{microtype}
\usepackage{graphicx}
\graphicspath{{media/}}
\usepackage{wrapfig}
\usepackage{amsmath}
\usepackage{caption}
\usepackage{xcolor}
\usepackage[most]{tcolorbox}
\usepackage{enumitem}

\usepackage[numbers,sort&compress]{natbib}

\usepackage[hidelinks]{hyperref}

\title{\textnormal{Componentization: Decomposing Monolithic LLM Responses into Manipulable Semantic Units}}

\author{
  Ryan Lingo \\
  Honda Research Institute, USA, Inc. \\
  \texttt{ryan\_lingo@honda-ri.com} \\
  \And
  Rajeev Chhajer \\
  Honda Research Institute, USA, Inc. \\
  \texttt{rajeev\_chhajer@honda-ri.com} \\
  \AND
  Martin Arroyo \\
  Honda Research Institute, USA, Inc. \\
  \texttt{martin\_arroyo@honda-ri.com} \\
  \And
  Luka Brkljacic \\
  Honda Research Institute, USA, Inc. \\
  \texttt{luka\_brkljacic@honda-ri.com} \\
  \AND
  Ben Davis \\
  Honda Research Institute, USA, Inc. \\
  \texttt{bdavis@honda-ri.com} \\
  \And
  Nithin Santhanam \\
  Honda Research Institute, USA, Inc. \\
  \texttt{nithin\_santhanam@honda-ri.com} \\
}

\begin{document}
\maketitle

\begin{abstract}
Large Language Models (LLMs) often produce monolithic text that is hard to edit in parts, which can slow down collaborative workflows. We present \textit{componentization}, an approach that decomposes model outputs into modular, independently editable units while preserving context. We describe \textit{Modular and Adaptable Output Decomposition (MAOD)}, which segments responses into coherent components and maintains links among them, and we outline the \textit{Component-Based Response Architecture (CBRA)} as one way to implement this idea. Our reference prototype, MAODchat, uses a microservices design with state-machine-based decomposition agents, vendor-agnostic model adapters, and real-time component manipulation with recomposition.

In an exploratory study with four participants from academic, engineering, and product roles, we observed that component-level editing aligned with several common workflows and enabled iterative refinement and selective reuse. Participants also mentioned possible team workflows. Our contributions are: (1) a definition of componentization for transforming monolithic outputs into manipulable units, (2) CBRA and MAODchat as a prototype architecture, (3) preliminary observations from a small user study, (4) MAOD as an algorithmic sketch for semantic segmentation, and (5) example Agent-to-Agent protocols for automated decomposition. We view componentization as a promising direction for turning passive text consumption into more active, component-level collaboration.
\end{abstract}


\section{Introduction}

Large Language Models (LLMs) are now common in knowledge work and creative tasks. People use them to draft prose, explain or synthesize code, sketch multi-step plans, and generate structured text like tables or JSON. Yet across these use cases, collaboration is often hampered by a core architectural limitation: LLMs return a single monolithic block. Whether the output is text, source code, a stepwise plan, or a structured reply, its all-or-nothing form resists targeted edits.

This monolithic structure creates friction. To fix a small part, users typically face two inefficient paths, a dilemma we call the \emph{Copy–Paste Problem} (Figure~\ref{fig:copy_paste}). One path is to copy the whole response into an external editor, severing it from the conversational context and foregoing further AI assistance \cite{Laban2024}. The other is iterative re-prompting, which risks overwriting good sections while chasing one local change and can make the conversation unwieldy \cite{Cai2024}. These patterns slow revision and make fine-grained collaboration harder, especially on complex tasks.

\begin{figure}[!t]
\centering
\begin{minipage}{\textwidth}
\begin{tcolorbox}[colback=gray!5!white,colframe=gray!75!black,title=The Copy–Paste Problem: Typical Workflows Across Output Types]
\small
\textbf{Scenario A (email):} LLM returns a 5-paragraph update. User needs only to adjust tone in paragraph 3.\\[0.25em]
\textbf{Current options:}
\begin{enumerate}[leftmargin=*,nosep]
\item Copy everything to an editor, edit paragraph 3, lose the link to the chat context and easy re-use.
\item Re-prompt ``make paragraph 3 more formal,'' risk unintended changes in paragraphs 1, 2, 4, and 5; repeat several times.
\end{enumerate}
\textbf{Outcome:} Manual effort or unwanted global changes.\\[0.6em]
\textbf{Scenario B (code / plan):} LLM returns a Python module or a 7-step plan. User needs to modify one function or step 4.\\[0.25em]
\textbf{Current options:}
\begin{enumerate}[leftmargin=*,nosep]
\item Paste into an IDE or doc, edit locally, break provenance with the conversation.
\item Re-prompt to change a single function/step, risk regressions elsewhere (imports, tests, later steps).
\end{enumerate}
\textbf{Outcome:} Context loss or cascade changes outside the target area.
\end{tcolorbox}
\end{minipage}
\caption{Two common failure paths when users try to adjust one part of a monolithic response. The issue appears across emails, code, plans, reports, and structured outputs.}
\label{fig:copy_paste}
\end{figure}

We propose \emph{componentization}, an output-centric approach that breaks an LLM response into discrete, semantically coherent units, called \emph{components}. Users can then edit, include or exclude, or regenerate these components in place.
 Components generalize beyond writing: for code they may be functions, classes, import blocks, or tests; for plans, individual steps or subgoals; for structured text, table rows/columns or JSON subtrees. This framing borrows from modular software principles such as separation of concerns and interchangeability \cite{Niu2025, Zhang2025}. Instead of treating a response as a final artifact, we treat it as a set of manipulable building blocks for user-directed composition.

To make this concrete, we outline the \emph{Component-Based Response Architecture (CBRA)}, which organizes generation, decomposition, manipulation, and recomposition as first-class stages, and we provide a reference prototype, \emph{MAODchat}, that illustrates the approach. While our prototype focuses on text-like outputs (natural language, code, pseudo-code, and structured text), the architecture is content-type-agnostic when outputs can be represented as typed components with links.

Our contributions are:
\begin{enumerate}
    \item \textbf{Conceptualization:} A definition of \emph{componentization} for turning monolithic LLM outputs, including prose, code, plans, and structured text, into manipulable units.
    \item \textbf{Architecture:} \emph{CBRA} and a \emph{MAODchat} prototype that illustrate a microservices-style design with vendor-agnostic model integration.
    \item \textbf{User Study:} Preliminary qualitative observations (n=4) suggesting that component-level editing aligns with several real workflows beyond writing.
    \item \textbf{Method:} \emph{MAOD}, an algorithmic sketch for semantic segmentation that preserves relationships among components.
    \item \textbf{Protocols:} Example Agent-to-Agent patterns for automated decomposition in multi-agent systems.
\end{enumerate}

\section{Related Work}
LLMs are reshaping the landscape of Human-Computer Interaction (HCI). While these models offer powerful generative capabilities, their integration into human workflows is still an active area of research. This review examines existing literature across four key domains: the nature of human-AI collaboration, the paradigm of prompt engineering, emerging techniques for direct output manipulation, and the future of multi-agent systems.

\subsection{Human-AI Interaction and Collaboration}
Recent HCI research has focused heavily on understanding how humans and AI collaborate, particularly in creative and professional tasks like writing \cite{Heyman2024, Zhang2025}. Studies of real-world use have identified various prototypical behaviors in LLM-assisted writing, revealing a complex interplay between the user and the model \cite{Mysore2025}. Prior work suggests that effective human–AI collaboration can improve written outputs in some settings \cite{Laban2024, Chen2025}. However, this potential is often constrained by existing interfaces, which can lack the affordances needed for deep collaboration and unstructured problem-solving \cite{Subramonyam2024}.

A primary challenge lies in the design of the interaction itself. Some researchers propose flipping the conventional dynamic, where the LLM asks questions to guide the human writer, thereby enhancing the collaborative experience \cite{Chen2025, Chin2025}. Others highlight the need for better user feedback mechanisms within the interface to guide the model's output more effectively \cite{Zhang2025, Reza2025}. The way users conceptualize and anthropomorphize LLMs significantly impacts the nature of the interaction and suggests the need for interfaces that clarify the AI's role as a tool \cite{Subramonyam2024, Reza2025}. This body of work points to a need for interfaces that move beyond the simple chat interface to support a richer, more controlled collaborative process \cite{Laban2024, Subramonyam2024}. While foundational models for human-AI collaboration have been proposed, our work offers a concrete architectural pattern that may help realize these models in practice.

\subsection{Prompt Engineering as the Dominant Paradigm}
Currently, the primary method for controlling and directing LLMs is prompt engineering \cite{Cai2024, Subramonyam2024}. This input-side paradigm has evolved from simple instructions into a sophisticated discipline \cite{Cai2024, Subramonyam2024}. The complexity of prompting has grown, leading to taxonomies of prompt modifiers for specialized tasks like image generation and even collaborative platforms where users can share and reference effective prompts for programming tasks \cite{Feng2024, Shen2025}.

However, the reliance on prompt engineering as the sole interaction method presents limitations \cite{Cai2024, Subramonyam2024}. It places much of the burden of content specification and refinement on the user before generation \cite{Cai2024, Subramonyam2024}. If the model's output is partially correct but flawed in one area, the user is forced to either perform manual edits outside the system or restart the prompting process \cite{Laban2024, Ugare2025}. This risks the loss of the valuable portions of the previous response \cite{Ugare2025}. While powerful, prompt engineering is an input-focused solution \cite{Cai2024, Chen2025, Shen2025}. Our Component-Based Response Architecture (CBRA) takes a step toward an output-focused solution for post-generation refinement.

\subsection{Direct Output Manipulation and Model Editing}
Recognizing the limitations of input-only interaction, a growing body of research is exploring methods for the direct manipulation and editing of LLM outputs \cite{Laban2024, Mysore2025, Cai2024, Subramonyam2024, Ugare2025}. The goal is to move beyond the chat interface toward systems that allow for more granular, verifiable, and structured control over the generated text \cite{Laban2024, Ugare2025}.

Several approaches are emerging. Some focus on creating interfaces for the direct manipulation of language models, allowing users to interact with the output in a more tangible way \cite{Laban2024, Zhang2025, Reza2025}. Others develop human-in-the-loop systems that enable users to augment and rewrite model outputs for tasks like query generation \cite{Chen2025, Ugare2025, Feng2024}. A more technical line of research investigates methods for ``model editing,'' which aim to distill edits from a small number of examples to update the model's behavior \cite{Mysore2025, Chakrabarty2025}.

This research underscores the need for output-level control \cite{Laban2024, Ugare2025}. However, these methods often require specialized interfaces or focus on altering the model's internal state \cite{Laban2024, Mysore2025, Cai2024, Ugare2025}. CBRA contributes to this area by proposing an architectural pattern that externalizes the editing process. Instead of modifying the model, it decomposes the output into a set of user-controllable components. This provides a simple, intuitive, and model-agnostic paradigm for structured editing, for which we offer preliminary observations from a small user study showing alignment with several natural workflows across professional contexts.

\subsection{The Future of LLM-Based Agents and Systems}
Looking forward, the field is rapidly moving toward the development of LLM-based agents \cite{Niu2025}. These are autonomous systems that can reason, plan, and execute tasks \cite{Cai2024, Chin2025, Niu2025}. As these foundation models become more capable and integrated into complex workflows, the HCI challenges related to control, transparency, and collaboration are likely to intensify \cite{Laban2024, Cai2024, Subramonyam2024, Reza2025}. The development of robust communication protocols and interaction patterns for these future agentic systems is a critical area of research \cite{Niu2025, Subramonyam2024}. The Agent-to-Agent (A2A) protocol implemented in our MAODchat system represents an initial step in this direction. It provides a reference for how specialized agents can communicate to perform sub-tasks within a larger human-AI collaborative system \cite{Niu2025, Feng2024, Liu2025, Subramonyam2024}.

\section{Conceptual Framework: The Component-Based Response Architecture (CBRA)}
The Component-Based Response Architecture is built upon three core principles that together support a move from monolithic generation toward modular composition. These principles define a structured workflow that begins with the deconstruction of an AI's output and culminates in a user-directed final artifact.

\begin{figure}[!htb]
\centering
\includegraphics[width=\textwidth]{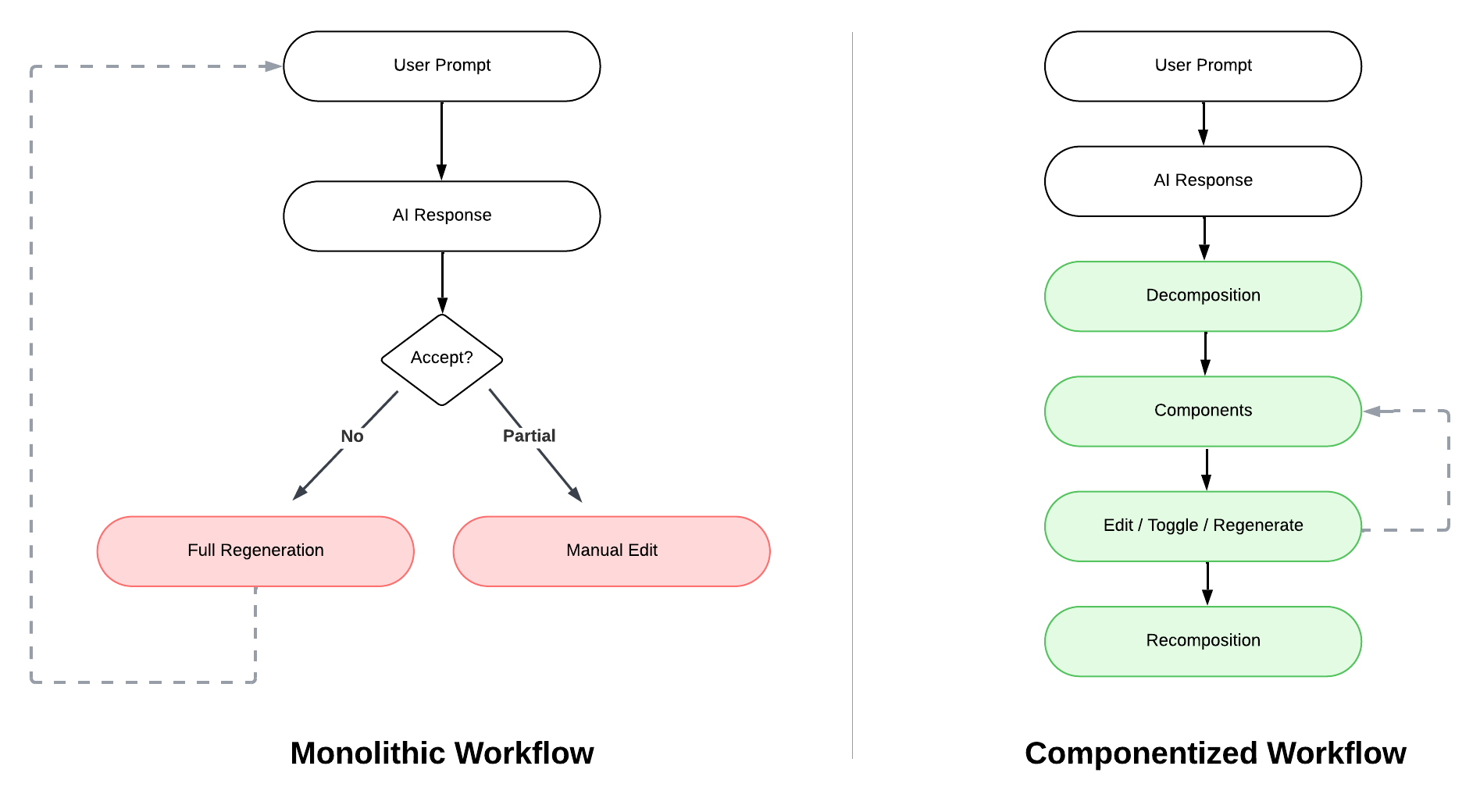}
\caption{Monolithic workflow versus componentized workflow. The componentized path enables edit, toggle, and
regenerate at the component level, then recomposes the final artifact.}
\label{fig:workflow_comparison}
\end{figure}

\subsection{Principle I: Modular and Adaptable Output Decomposition (MAOD)}
The foundational principle of CBRA is Modular and Adaptable Output Decomposition (MAOD). This is the process of transforming a single, monolithic LLM response into a structured collection of discrete, semantically-coherent components. It is critical to distinguish this from simple text splitting by paragraphs or sentences. MAOD is a semantic process that identifies the logical, functional units of a response based on its context and intent. For example, an email generated by an LLM would not be split into arbitrary paragraphs but would be decomposed into its constituent parts: a Subject, Greeting, Body Paragraphs, Closing, and Signature.

Formally, a monolithic response $R$ is passed through $f_{\text{maod}}$ to produce a set $C=\{c_1, c_2, \ldots, c_n\}$. Each $c_i$ stores content and metadata. The transformation yields a machine-readable and user-manipulable structure, implemented as a \texttt{DecomposedResponse}. A minimal schema is shown in Table~\ref{tab:schema}.

\begin{table}[t]
\centering
\caption{Minimal schema for a decomposed response.}
\label{tab:schema}
\begin{tabular}{llp{8cm}}
\toprule
Field & Type & Description \\
\midrule
\texttt{id} & string & Stable component identifier \\
\texttt{type} & enum & Component class (Heading, Paragraph, List, Code, Citation) \\
\texttt{content} & string & Component text payload \\
\texttt{meta} & map & Metadata (level, role, style) \\
\texttt{includes} & bool & Whether the component is selected for recomposition \\
\texttt{links} & list & Inter-component relations (for example, \texttt{belongs\_to: c1}) \\
\bottomrule
\end{tabular}
\end{table}

\paragraph{MAOD procedure.}
Given response $R$, MAOD performs:
(1) Parse: detect blocks, lists, code, and citations,
(2) Segment: propose spans using rhetorical and structural cues,
(3) Classify: assign a component type and attach metadata,
(4) Link: infer relations (for example, a paragraph belongs to a section),
(5) Validate: check constraints such as no empty components and acyclic links,
(6) Export: return a \texttt{DecomposedResponse} for manipulation.

A small running example clarifies the representation:
\begin{verbatim}
[
  { "id": "c1", "type": "Subject", "content": "Project update" },
  { "id": "c2", "type": "Greeting", "content": "Hi team," },
  { "id": "c3", "type": "Paragraph", "content": "We shipped v1.2 today...", 
    "links": ["c1"] }
]
\end{verbatim}

\subsection{Principle II: User-Driven Component Manipulation}
Once decomposition is complete, the components, not the original monolithic response, become the primary objects of interaction. CBRA empowers the user with a set of granular operations to directly control these components. This changes the user's role from a passive prompter to an active composer. The core affordances for manipulation include:

\begin{itemize}
    \item \textbf{Edit:} Users can modify the content of any individual component in-place, correcting errors, refining tone, or adding information without affecting any other part of the response.
    \item \textbf{Select/Toggle:} Users can choose to include or exclude specific components from the final output. This allows for the easy removal of irrelevant sections or the selection of only the most critical pieces of information.
    \item \textbf{Regenerate:} Users can generate a new version of a component, refreshing its content without affecting the overall structure of the final document.
\end{itemize}

These operations add direct control that most chat UIs lack, fostering a more dynamic and collaborative human-AI partnership \cite{Laban2024, Chen2025, Shen2025}.

\subsection{Principle III: Dynamic and Resilient Recomposition}
The final principle is the generation of the output through Dynamic and Resilient Recomposition. The final document is not static. It is a dynamic composition that is a direct function of the user's manipulations on the component set.

This approach can add \textit{resilience}. In a monolithic system, a single flaw often requires a complete regeneration of the response, which may introduce new, unforeseen errors. Under CBRA, a flaw in one component is localized. It can be edited or removed independently, preserving the integrity of the remaining valuable components. This decoupling of component quality helps avoid the ``catastrophic regeneration'' failures sometimes seen in monolithic workflows. This entire cycle is visually represented in our system's Four-Column Interface:

\begin{figure}[!htbp]
\centering
\includegraphics[width=\textwidth]{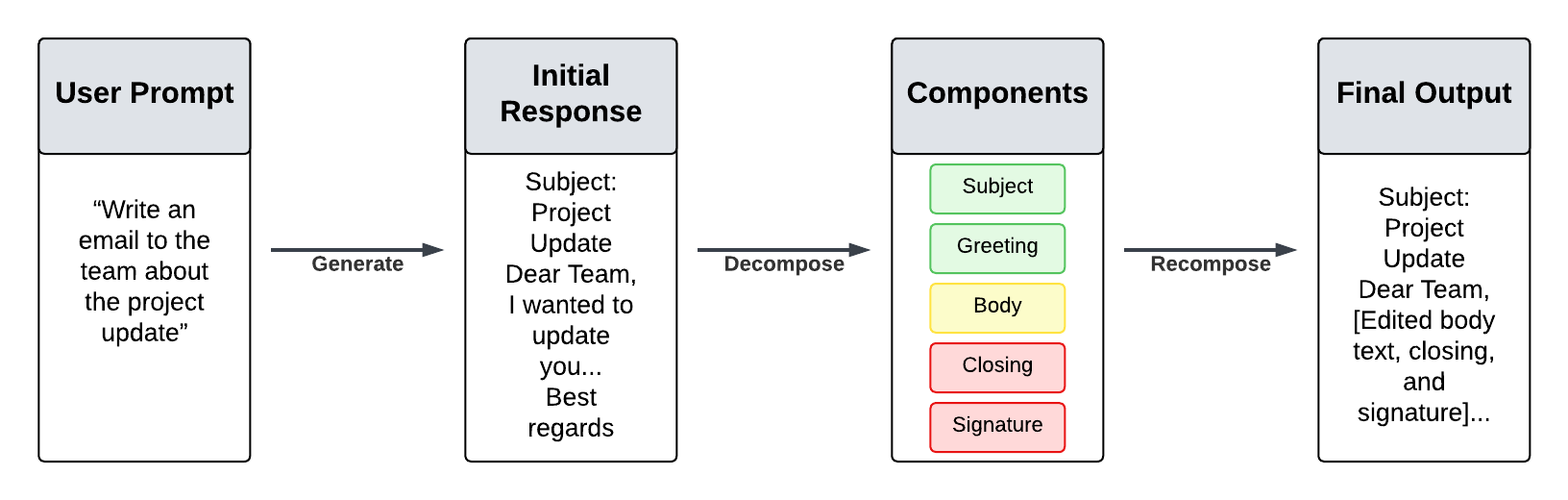}
\caption{The Four-Column Interface implementing CBRA principles. Users input prompts (Column 1) which generate initial AI responses (Column 2). These responses undergo automatic decomposition into semantic components (Column 3), where users can perform granular operations: editing individual components, toggling inclusion, and regenerating. The final output (Column 4) dynamically reflects user manipulations, providing immediate visual feedback. This interface moves beyond the traditional linear chat interface into an interactive composition workspace.}
\label{fig:four_column_interface}
\end{figure}

\section{MAODchat: A Reference Implementation}
\label{sec:maodchat_implementation}
To examine the principles of the Component-Based Response Architecture, we developed MAODchat, a full-stack reference implementation. This section details the system's architecture, core components, and the key technical choices that enable its functionality.

\subsection{High-Level System Design}
\label{subsec:high_level_design}
MAODchat is implemented using a Service-Oriented Architecture (SOA) composed of five interconnected microservices. This design was chosen over a monolith to provide independent deployment, fault isolation, and specialized scaling capabilities for each service \cite{Niu2025}. The architecture consists of a Flask-based Frontend, a FastAPI Backend, a FastAPI MAOD Agent for decomposition, and a PostgreSQL database for persistence, all orchestrated by a Caddy reverse proxy.

This microservices approach allows for tailored resource management. The stateless Backend is designed for horizontal scaling to handle concurrent user requests, while the compute-intensive MAOD Agent is suited for vertical scaling to manage complex decomposition tasks. The database acts as the central, persistent state manager, with services communicating via well-defined REST APIs.

\begin{figure}[ht]
\centering
\includegraphics[width=0.9\textwidth]{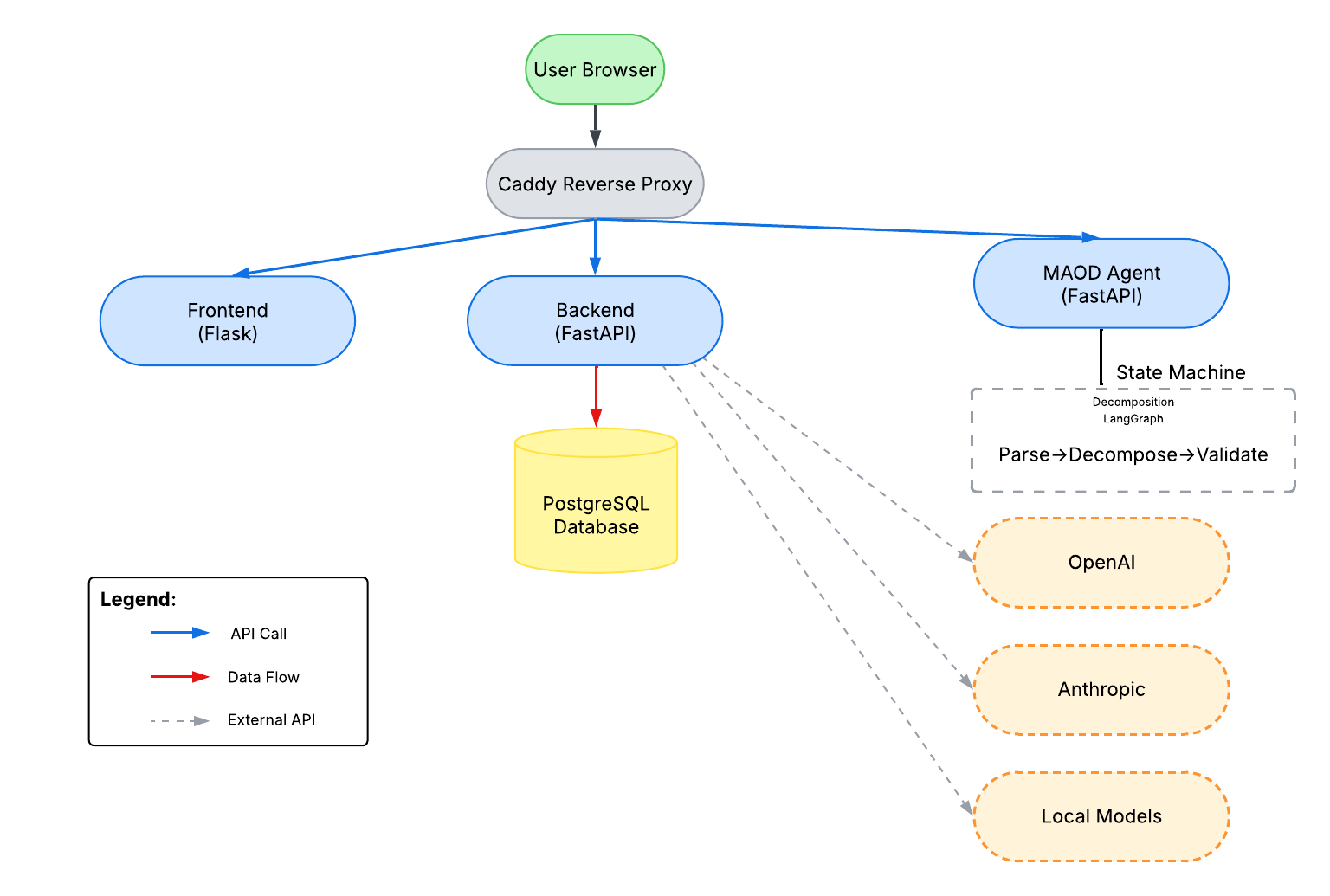}
\caption{MAODchat microservices architecture. The system employs a Service-Oriented Architecture with five core services orchestrated by a Caddy reverse proxy. The Backend service manages orchestration and session state while maintaining vendor-agnostic connections to LLM providers through a Dynamic Model Factory pattern. The MAOD Agent implements decomposition logic via an internal state machine (Parse→Decompose→Validate) using LangGraph. Services communicate through REST APIs and the specialized Agent-to-Agent (A2A) protocol. PostgreSQL provides persistent state management across user sessions.}
\label{fig:system_architecture}
\end{figure}

\subsection{Core Components and Algorithms}
\label{subsec:core_components_algorithms}

\subsubsection{The MAOD Agent}
\label{subsubsec:mod_agent}
The MAOD Agent is the specialized service responsible for executing the Modular and Adaptable Output Decomposition (MAOD) principle. It receives a monolithic text response from the Backend and transforms it into a structured format. To manage the multi-step decomposition process, the agent utilizes LangGraph to implement a state machine-based workflow. The final output is a structured, type-safe object defined by a Pydantic \texttt{DecomposedResponse} model, which provides a reliable data contract with the other services in the system. The agent is designed for compliance with the Agent-to-Agent (A2A) communication protocol.

\subsubsection{The Backend Service}
\label{subsubsec:backend_service}
The Backend service acts as the central orchestrator, managing business logic, user sessions, and interactions with external LLM providers. A key feature is a Dynamic Model Factory Pattern, which uses Python's reflection capabilities (\texttt{importlib}) to instantiate LLM clients at runtime based on user selection. This reduces hard-coded model implementations and aims to support a vendor-agnostic system.

Conversation state management is handled using LangGraph, which models the conversation flow as a state graph. The state at any given time $t$ can be represented as $S(t) = \{M(t), C(t), E(t), \theta(t)\}$, where $M$ is the message history, $C$ is the context, $E$ are user events, and $\theta$ are model parameters. This state is persisted using an \texttt{AsyncPostgresSaver} checkpointer, helping maintain long-term, stateful conversations.

\subsubsection{The Frontend Service}
\label{subsubsec:frontend_service}
The Frontend is a lightweight Flask application responsible for rendering the user interface and managing client-side state. It implements the four-column interface described in the conceptual framework. Using vanilla JavaScript, the frontend is responsible for the real-time state management of components, tracking user edits, selections, and regenerating. It dynamically generates the final composed output in the fourth column based on the current state of the components in the third column, providing immediate visual feedback to the user.

\subsection{Key Technical Innovations}
\label{subsec:key_innovations}
Two technical elements in MAODchat are central to its implementation of CBRA.

\subsubsection{Agent-to-Agent (A2A) Protocol for Decomposition}
\label{subsubsec:a2a_protocol}
MAODchat features a working implementation of an Agent-to-Agent (A2A) protocol designed specifically for the task of response decomposition. This protocol standardizes communication between the Backend and the MAOD Agent, enabling type-safe message passing and clear task delegation. This architectural pattern is designed to be extensible, laying the groundwork for potential integrations with additional specialized agents, such as fact-checking or formatting agents, in a multi-agent workflow.

\subsubsection{Vendor-Agnostic Model Abstraction}
\label{subsubsec:vendor_abstraction}
To address the challenge of incompatible APIs and configuration formats across different LLM providers, MAODchat implements a vendor abstraction layer. This layer uses a centralized enum, \texttt{VendorMetadata}, to map provider-specific details, such as parameter names (\texttt{model\_name\_key}, \texttt{temperature\_key}) and module paths, to a standardized internal representation. This approach, combined with the model factory, aims to decouple the core application logic from any specific LLM provider, facilitating integration of new models and vendors without code changes to the core system.

\section{Analysis and Discussion}
\label{sec:analysis_discussion}
The implementation of MAODchat provides a practical foundation for analyzing the Component-Based Response Architecture. This section discusses the direct impacts of CBRA on user workflow and system properties, while also acknowledging the inherent limitations of the approach.

\subsection{Impact on User Agency and Workflow}
\label{subsec:user_agency}
CBRA, as implemented in MAODchat, can enhance user agency by shifting the interaction model from a simple request–response cycle toward a more co-creative partnership \cite{Chen2025, Reza2025}. Our user study suggested that this modularity can enable more advanced workflows. For instance, participants described a ``scaffolding'' technique where users generate a complex report, then toggle off all components except the headings to review and rearrange the document's high-level structure. As one participant noted, this approach ``matched well with [the] process of building outlines, then revising sections iteratively.'' This level of granular control is difficult to achieve in monolithic chat interfaces and can help users shape the final output with greater precision.

\subsection{System Resilience and Error Handling}
\label{subsec:resilience_error_handling}
The architecture exhibits resilience on two distinct levels: content and system architecture.

\begin{itemize}
    \item \textbf{Content Resilience:} The core principle of decomposition can provide content-level resilience. Errors or undesirable content generated by the LLM are localized to individual components. This allows for surgical correction without risking a ``catastrophic regeneration,'' where a request to fix one problem results in a new, different response with new flaws.
    
    \item \textbf{Architectural Resilience:} The microservices design supports graceful degradation in our prototype. For example, if the MAOD Agent service fails, the system can fall back to presenting a monolithic response, ensuring that the application remains functional. Furthermore, the implementation of a custom exception hierarchy (e.g., \texttt{ModelInitializationError}, \texttt{FileProcessingError}) provides context-aware error handling, which can improve system robustness and maintainability.
\end{itemize}

\subsection{Scalability and Performance}
\label{subsec:scalability_performance}
The system is designed for scalability. The stateless nature of the Backend and MAOD Agent services allows for straightforward horizontal scaling behind a load balancer to handle a high volume of concurrent users. Efficient database connection pooling further helps the persistence layer manage increased load effectively.

A key consideration is the performance trade-off. The decomposition step necessarily introduces a small amount of latency compared to directly streaming a response to the user. This is mitigated through several optimizations. Model instance caching avoids the costly re-initialization of LLM clients for repeated requests. Additionally, all database operations are performed asynchronously, preventing I/O-bound tasks from blocking the server and helping the system remain responsive.

\subsection{Limitations}
\label{subsec:limitations}
For a balanced analysis, it is important to acknowledge the limitations of the current CBRA implementation.

\begin{itemize}
    \item \textbf{Decomposition Overhead:} As noted, the MAOD process adds a latency penalty. For applications where real-time streaming is paramount, this trade-off may not be acceptable.
    
    \item \textbf{Semantic Accuracy:} The effectiveness of the entire user experience hinges on the quality of the decomposition performed by the MAOD Agent. A poorly segmented or semantically inaccurate breakdown could frustrate the user and make the editing process more difficult than in a monolithic system. The system's utility is therefore highly dependent on the intelligence of its decomposition agent.
    
    \item \textbf{Component Interdependence:} The current model treats all components as fully independent. However, many documents have logical dependencies; for example, the content of a conclusion should reflect the points made in the introduction. The system does not currently detect these interdependencies or assist the user in maintaining coherence across components during editing.
\end{itemize}

\section{User Validation Study}
\label{sec:user_validation}
To assess the practical utility of the Component-Based Response Architecture and explore its real-world applicability, we conducted semi-structured interviews with four participants from diverse professional backgrounds. This section presents the methodology, key findings, and implications from these user sessions.

\subsection{Study Methodology}
\label{subsec:study_methodology}
We recruited four participants with varied expertise: an academic researcher (Participant A), a product manager with HCI background (Participant B), and two software engineers (Participants C and D). Each participant engaged in a 45–60 minute session consisting of hands-on interaction with MAODchat followed by semi-structured interviews. Participants were asked to complete both prescribed tasks (email drafting, code generation) and self-selected tasks relevant to their workflows. Sessions were conducted remotely via video conferencing, with participants sharing their screens while thinking aloud.

\begin{center}
\begin{tabular}{lll}
\toprule
ID & Role & Tasks performed \\
\midrule
A & Academic researcher & Outline creation, section rewrite \\
B & Product manager (HCI) & Slide text structuring and trimming \\
C & Software engineer & Code explanation and refactor \\
D & Software engineer & Config transformation \\
\bottomrule
\end{tabular}
\captionof{table}{Participant profiles and primary tasks.}
\label{tab:participants}
\end{center}

\subsection{Key Findings}
\label{subsec:key_findings}

\subsubsection{Validation of Core Decomposition Value}
Participants generally recognized value in decomposition for their workflows. Participant A, working primarily with academic writing, noted that the decomposition ``matched well with [the] process of building outlines, then revising sections iteratively.'' This observation is consistent with our assumption that componentization aligns with natural writing processes. Participant B highlighted the utility for PowerPoint workflows, stating that users ``often want to break text into slides or vice versa, making decomposition highly relevant.''

The ability to remove unnecessary content emerged as a recurring benefit. Multiple participants independently identified the removal of ``fluff blocks'' (introductions and conclusions) as addressing a common frustration with AI-generated content. As Participant B noted, this ``mirrors real-world student and essay workflows'' where such content is often superfluous.

\subsubsection{Interface Design and Mental Models}
A key finding concerned user mental models and expectations. Participants often brought ChatGPT-based expectations to the interface, creating friction when MAODchat deviated from these norms. The distinction between ``Edit'' and ``Regenerate'' functions proved confusing, with Participant C expecting ``Edit to allow inline adjustments without triggering a full re-prompt.'' This confusion was echoed across multiple sessions, suggesting a mismatch between our terminology and established user expectations.

The four-column layout, while conceptually clear to the research team, created unexpected cognitive load. Participant C expressed a preference for ``a single top-to-bottom flow like ChatGPT,'' indicating that our departure from established patterns may have introduced unnecessary complexity.

\subsubsection{Technical Constraints and Performance}
Performance emerged as a concern, particularly for complex use cases. Participant D's attempt to convert Docker Compose to Helm configurations resulted in system failures, likely due to context window limitations. This highlighted a gap between user expectations for real-world applications and current system capabilities. However, when constrained to appropriately scoped tasks (such as simple Python scripts), the system performed well and showed clear value.

Formatting preservation proved problematic across sessions. Participant B noticed that ``markdown, numbered lists'' were sometimes lost during decomposition, describing this as ``disruptive.'' This technical limitation affected the perceived utility of the system for structured documents.

\subsubsection{Collaboration Potential}
Participants also envisioned collaborative possibilities. Several independently mentioned team-based workflows enabled by componentization. Participant A drew a parallel to ``GitHub for papers,'' describing a system where ``contributions could be controlled and reintegrated cleanly without loss of ownership.'' 

Participant B elaborated on a workflow where ``a manager decomposes a project and distributes sections to teammates to edit independently before reintegrating.'' This recurring recognition suggests that componentization could support multi-user workflows.

\subsection{Design Implications}
\label{subsec:design_implications}
The validation study revealed several design implications:

\textbf{Terminology Alignment}: The confusion around ``Edit'' versus ``Regenerate'' suggests aligning terminology with established patterns. Participants suggested alternatives like ``Manual Edit'' and ``Reprompt'' that may better match user expectations.

\textbf{Formatting Fidelity}: The loss of formatting during decomposition is a barrier to adoption. Maintaining markdown, lists, and other structural elements appears essential for practical utility.

\textbf{Context Management}: The system should better communicate its context limitations and guide users toward appropriately scoped tasks. Participant A's suggestion of ``scaffolded prompts for research papers'' points toward a solution that provides structure while managing expectations.

\textbf{Progressive Disclosure}: The four-column interface may benefit from progressive disclosure that initially presents a simpler view closer to familiar chat interfaces, with advanced features revealed as users gain expertise.

\subsection{Limitations of the Study}
\label{subsec:study_limitations}
This validation study has several limitations. The small sample size (n=4) limits generalizability, though the diversity of backgrounds provided qualitative insights. The remote, screen-sharing format may have introduced technical complications that influenced user perception. Additionally, the prototype's technical limitations (particularly around complex inputs) prevented full exploration of the system's potential.

Despite these limitations, the study surfaced consistent signals of value for the componentization concept while identifying specific areas for refinement.

\section{Future Work and Research Directions}
\label{sec:future_work}
The Component-Based Response Architecture, as realized in MAODchat, opens up several directions for future research and development. The following areas aim to address current limitations and expand upon the core contributions of this work.

\subsection{Advanced Decomposition Models}
\label{subsec:advanced_decomposition}
The utility of the CBRA paradigm depends on the semantic accuracy of the initial decomposition. A direction for future work is to enhance the MAOD Agent by exploring models specialized for semantic segmentation of text. Such models could be optimized for recognizing document structures, identifying functional units, and providing more consistent component breakdowns.

\subsection{Expanded Usability Studies and Workflow Analysis}
\label{subsec:usability_analysis}
Future studies could quantify impact through:
\begin{itemize}
    \item Larger-scale quantitative studies measuring task time, error rates, and satisfaction compared to monolithic interfaces
    \item Longitudinal studies examining how users adapt workflows over extended use
    \item Domain-specific evaluations in academic writing, software development, and business communication
    \item A/B tests of interface variations, particularly terminology (``Edit'' vs ``Regenerate'') and layout preferences
\end{itemize}
Given participants' interest in team scenarios, multi-user workflows merit focused investigation.

\subsection{Expansion of Agentic Workflows}
\label{subsec:agentic_workflows}
The A2A protocol in MAODchat serves as a proof of concept for inter-agent communication. Future work could expand this into a richer ecosystem of specialized agents, creating pipelines such as: Decomposition → Fact Verification → Citation Checking → Formatting, before presentation to the user.

\subsection{Automated Component Coherence}
\label{subsec:automated_coherence}
A longer-term challenge is handling inter-component dependencies. We propose investigating mechanisms for automated component coherence, where the system can detect logical dependencies and either flag inconsistencies or suggest coordinated edits across components.

\subsection{Collaborative Team-Based Workflows}
\label{subsec:collaborative_workflows}
The component-based structure is a natural foundation for multi-user environments. Future development could explore assigning components to specific users, tracking per-component changes, and merging contributions. This would move the system toward a collaborative hub for team-based content creation.

\section{Conclusion}
Many LLM interfaces still present long, single blocks of text. That shape makes fine-grained editing and collaboration harder than it needs to be. We set out a different pattern. Treat an output as a set of parts that can be acted on directly.

CBRA and the MAODchat prototype show one way to do this. The system decomposes a response into typed components with stable identifiers and relations, exposes three operations (Inline Edit, Toggle, Rewrite (model)), and then recomposes a final artifact. This shifts the unit of interaction from the whole document to the specific piece a person cares about, which is how people already think when they write or review.

Our small, exploratory study suggests that this framing maps to real workflows. Participants used component-level editing to keep what worked, trim what did not, and try targeted rewrites without losing good content elsewhere. They also imagined team workflows that divide and reintegrate work at the component level. These are early signals, not general claims, but they point in a clear direction.

We see componentization as a big idea that is simple to state and widely useful. Treat model outputs as structured objects rather than strings. When parts carry stable IDs, types, and relations, local changes become first-class operations. That enables practical behaviors like diff and merge, provenance, partial review, and permissions. These same ideas helped software collaboration scale, and they are a good fit for AI-assisted output.

The pattern likely extends beyond writing. Code refactors, data analysis notebooks, and UI flows all benefit when a user can change one part without disturbing the rest. In short, treating outputs as objects rather than blobs offers a direct path to more controllable, auditable, and collaborative AI tools.

\bibliographystyle{apalike}
\bibliography{my_references}

\end{document}